\documentclass[12pt,a4paper]{scrartcl}

\usepackage{graphicx}
\usepackage{amsmath}
\usepackage{amssymb}
\usepackage{xcolor}
\usepackage{textcomp} 
\usepackage{ulem} 
\usepackage[auth-sc-lg, affil-it]{authblk} 
\usepackage{mathbbol} 

\usepackage[implicit=true,colorlinks=true,linkcolor=blue,citecolor=blue,urlcolor=blue]{hyperref}
\usepackage[numbers,sort&compress]{natbib}

\DeclareMathOperator{\Imag}{\mathrm{Im}}
\newcommand{\units}[1]{\ensuremath{\,\mathrm{#1}}}
\newcommand{\fracd}[3][...]{\displaystyle\frac{d^{#1}#2}{d#3^{#1}}}
\newcommand{\fracp}[3][...]{\displaystyle\frac{\partial^{#1}#2}{\partial#3^{#1}}}
\newcommand{\SE}{Schr\"odinger equation}
\newcommand{\up}{\mathrm{l}}
\newcommand{\lw}{\mathrm{r}}

\newcommand{\ii}{\mathrm{i}}
\newcommand{\ee}{\mathrm{e}}

\definecolor{uniorange}{RGB}{250, 100, 55}
\definecolor{darkgreen}{rgb}{0.2,.7,0.2}

\usepackage{xcolor,cancel}

\begin{document}

\title{Revising de Broglie-Bohm trajectories' momentum distribution}

\author[1]{Dennis M. Heim\thanks{physics@d-heim.de}}
\affil[1]{Fraunhofer Institute for Industrial Mathematics (ITWM), 67663 Kaiserslautern, Germany}
\maketitle

\date{\today}

\begin{abstract}
    Initial momenta of de Broglie-Bohm trajectories generally do not obey quantum mechanical momentum distributions.
    The solution to this problem presented in the following leads to an extended hydrodynamic interpretation of quantum mechanics.
    To demonstrate the new formalism, it is applied to the double-slit experiment.
    From this example follows a proposal for experimental questioning of trajectory formulations of quantum mechanics.
\end{abstract}

\section{Introduction}\label{sec:intro}

That de Broglie-Bohm trajectories \cite{bohm:1952:1} can disagree with quantum mechanical momentum distributions is well known \cite{jung:2013,nauenberg:2014,bonilla:2020}.
One thinks e.g. of a particle that is initially described by a Gaussian wave function
\begin{align}
    \psi(x) = \frac{1}{\sqrt[4]{2\pi}\sqrt{\sigma}} \ee^{-\frac{x^2}{4 {\sigma}^2}}
    . \label{eq:gauss}
\end{align} 
In quantum mechanics, the probability to find this particle is also given by a Gaussian function, as is the probability of measuring its momentum.

In contrast, the de Broglie-Bohm theory states that the momentum of this particle is always fixed at the value zero.
Such a result is obtained because the de Broglie-Bohm theory defines the particle momentum
\begin{align}
    p \equiv {\hbar}\cdot \Imag\left( \frac{\partial\psi/\partial x}{\psi} \right)
    \label{eq:momentum:def}
\end{align}
via an imaginary part which is zero for the real wave function $\psi \equiv \psi(x)$ \eqref{eq:gauss}.

Further thought experiments of similar nature lead to the conclusion that in order to bring the de Broglie-Bohm theory in line with quantum mechanics, not only the initial positions but also the initial momenta should be chosen according to their quantum mechanical distribution.
We want to make this conclusion a precondition and additionally introduce the precondition that the continuity equation for the probability density remains valid.

These two conditions lead to an extended hydrodynamic interpretation of quantum mechanics, in which the intrinsic kinematics of a particle is considered in addition to the kinematics of the fluid \cite{madelung:1926} where it is diluted in.
We illustrate this interpretation by showing trajectories for a double-slit experiment.
This demonstration leads to a proposal for experimental questioning of trajectory formulations of quantum mechanics.

\section{Intrinsic particle kinematics}\label{sec:traj}

In order to correct de Broglie-Bohm trajectories by adapting their initial momenta to the quantum mechanical ones, we have to extend the momentum definition \eqref{eq:momentum:def}.
We do this by adding a function $f(x,t)$ which leads to a revised momentum field
\begin{align}
    p_\mathbf{r}(x,t) \equiv p_\mathbf{BB}(x,t) + f(x,t)
    , \label{eq:momentum:ansatz}
\end{align}
where we used the de Broglie-Bohm momentum field
\begin{align}
    p_\mathbf{BB}(x,t) \equiv {\hbar}\cdot \Imag\left[ \frac{\partial\psi(x,t)/\partial x}{\psi(x,t)} \right]
    .
\end{align}

To determine $f(x,t)$, we insert the definition \eqref{eq:momentum:ansatz} into the continuity equation
\begin{align}
    \frac{\partial}{\partial t}  \rho(x,t) + \frac{\partial }{\partial x} \left[ {\rho(x,t)\, \frac{p_\mathbf{r}(x,t)}{m}} \right] = 0
    , \label{eq:continuity}
\end{align}
where $\rho(x,t)=|\psi(x,t)|^2$ is the quantum mechanical probability density.
From the continuity equation \eqref{eq:continuity} follows that the additional function can be written as
\begin{align}
    f(x,t) = \frac{c(t)}{\rho(x,t)}
    .
\end{align}
By demanding the momentum field \eqref{eq:momentum:ansatz} to equal the initial momentum $p_0 \equiv p_\mathbf{r}(x_0, t_0)$ at the initial position $x_0$ and initial time $t_0$, we arrive at
\begin{align}
    c(t) = [p_0-p_\mathbf{BB}(x_0, t_0)]\, \rho(x_0, t)
    .
\end{align}

In summary, we obtain the revised trajectories $x\equiv x(t)$ by solving the differential equation
\begin{align}
    m \fracd[]{x}{t} = p_\mathbf{r}(x,t) = p_\mathbf{BB}(x,t) + [p_0-p_\mathbf{BB}(x_0, t_0)]\,
    \frac{\rho(x_0, t)}{\rho(x, t)}
    \label{eq:momentum:summary}
\end{align}
and the de Broglie-Bohm trajectories by solving
\begin{align}
    m \fracd[]{x}{t} = p_\mathbf{BB}(x,t)
    .\label{eq:dbb-dgl}
\end{align}
We choose the initial positions $x_0$ randomly according to the distribution $|\psi(x,t_0)|^2$ and the initial momenta $p_0$ randomly from the momentum representation of this distribution, which is the absolute square of the Fourier transformed wave function. Trajectories $p(t)$ in momentum space are obtained by inserting each position and time of the trajectory $x(t)$ into the right side of equation \eqref{eq:momentum:summary} or \eqref{eq:dbb-dgl} respectively.

The part after the $+$ sign in the differential equation \eqref{eq:momentum:summary} can be interpreted as intrinsic particle movement which is triggered by the particle's initial momentum $p_0$.
It adds to the movement $p_\mathbf{BB}(x,t)$ of the fluid \cite{madelung:1926} where it is diluted in.
The quantum mechanical continuity equation \eqref{eq:continuity} still holds.

\section{Double-slit experiment}\label{sec:double-slit}

\begin{figure}[t!]
\begin{center}
    \includegraphics[]{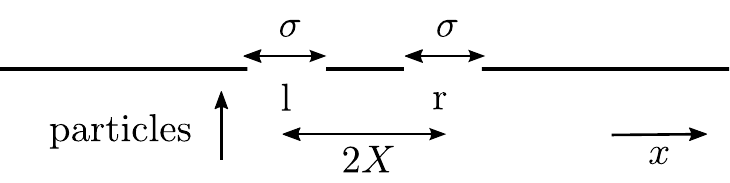}
\end{center}
    \caption{Double-slit experiment, where particles can pass two slits denoted by l and r with a distance $2X$. Each slit has a width $\sigma$.}
    \label{fig:setup}
\end{figure}

In the following, we calculate particle trajectories $x(t)$ for a double-slit experiment which is sketched in figure \ref{fig:setup}. For this purpose, we use the de Broglie-Bohm theory and the revised theory. The results are compared in position and in momentum space.

In quantum mechanics, the double-slit experiment is described by a superposition of two wave functions
\begin{align}
  {{\psi}}(x,t) = \frac{{\psi}_{\up}(x,t) + {\psi}_{\lw}(x,t)}{\sqrt{N}}
  , \label{eq:final-fuction-time}
\end{align}
which describes the possibility of finding particles along the $x$ axis at a time $t$ behind the slits in figure \ref{fig:setup}.
It consists of the sum of the wave functions ${\psi}_{\up}(x,t)$ passing the left and ${\psi}_{\lw}(x,t)$ passing the right slit, and the norm $N$.
    
For simplicity, we represent these wave functions by Gaussians
\begin{align}
	{\psi}_{\up,\lw}(x,t) =  {\frac{1}{\sqrt[4]{2\pi}\sqrt{\sigma+\ii {\hbar t}/{({2}m{\sigma}})}}} 
	 \exp  \left[ -\frac{(x\pm X )^2}{4{{\sigma}_{}}^2+2\ii {\hbar t}/m} \right] 
	 \label{eq:final-functions-single-time}
\end{align}
which solve the \SE{}
\begin{align}
	\ii \hbar \fracp[]{}{t} \psi = -\frac{\hbar^2}{2m} \fracp[2]{}{x} \psi
	.\label{eq:schroedinger:full}
\end{align}

\begin{figure}[t!]
	\begin{center}
		\includegraphics[width=15.0cm]{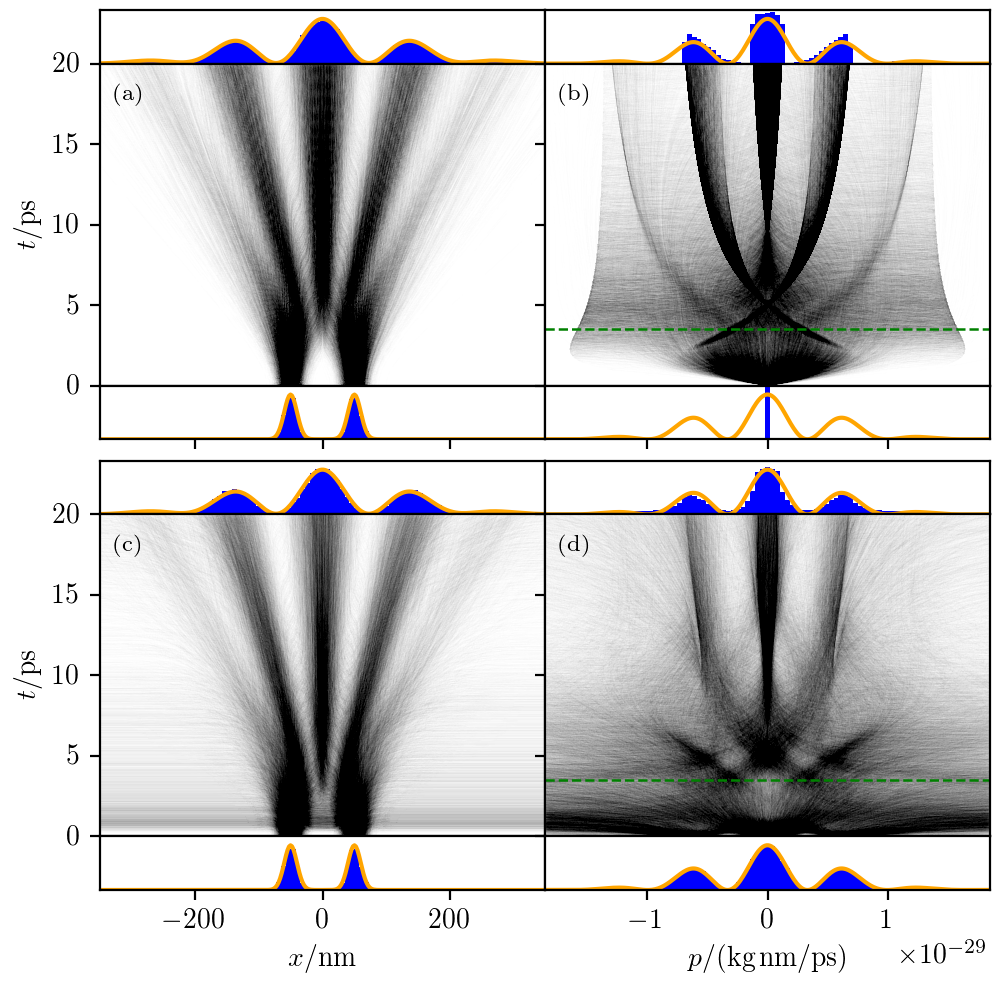}
		\caption{40000 trajectories are compared for a double-slit experiment with slit separation $2X$=100\units{nm} and slit width $\sigma$=10\units{nm}.
        In position space, histograms (blue) of the de Broglie-Bohm theory (a) as well as of the revised theory (c) match the quantum mechanical predictions (orange lines) at initial times and end times.
        In momentum space, however, the de Broglie-Bohm theory (b) mismatches the quantum mechanical initial and final distributions.
        In comparison, the revised theory (d) initially matches quantum mechanics perfectly but also deviates from the prediction at the end of the trajectories.
        In fact, quantum mechanics predicts the momentum distribution \eqref{eq:prob-density-momentum} here to be constant in time which is neither fulfilled in (b), nor in (d).
        We discuss particularly strong deviations, marked by dashed green lines at $t=3.5$\units{ps}, in section \ref{sec:conclusion}.
        }
		\label{fig:dslit}
	\end{center}
\end{figure}

Here, $2X$ is the slit separation and $\sigma$ describes how much the probability density expands in a single slit.
The norm of the wave function \eqref{eq:final-fuction-time} is then determined as $N=2+2\exp[-X^2/(2\sigma^2)]$.

We use this wave function in the differential equation \eqref{eq:dbb-dgl} to calculate de Broglie-Bohm trajectories and in equation \eqref{eq:momentum:summary} to calculate trajectories for the revised theory, respectively.
Figure~\ref{fig:dslit} shows these trajectories for parameters similar to the ones of the experiment performed by J\"onsson et al. \cite{joensson:1961}. The slit separation is taken as $2X$=100\units{nm} and the width as $\sigma$=10\units{nm}.
Every subfigure in figure~\ref{fig:dslit} consists of trajectories (center), whose initial values (bottom) and final values (top) are displayed as histograms (blue). Both histograms are compared to quantum mechanical predictions (orange lines).

As already reported by Bohm and Hiley~\cite{bohm:1975}, the de Broglie-Bohm trajectories in position space, which are displayed in figure \ref{fig:dslit}(a), are arranged along the valleys of an interference pattern.
For the histograms at initial times and at end times we can see that the quantum mechanical probability density $|\psi(x,t)|^2$ is matched.
The corresponding momenta of the de Broglie-Bohm trajectories, shown in figure \ref{fig:dslit}(b), initially violate the quantum mechanical probability density
\begin{align}
    \left|\widetilde{\psi}(p)\right|^2=\frac{\sqrt{{2}/{\pi}}/\sigma_p}{1+\exp(-2{\sigma_p}^2X^2/\hbar^2)}\exp\left(-\frac{p^2}{2{\sigma_p}^2}\right)\cos^2\left(X\frac{p}{\hbar}\right), \quad \sigma_p\equiv\frac{\hbar}{2\sigma}
    , \label{eq:prob-density-momentum}
\end{align}
but at final times the histograms are partly adapted to this density.

The revised trajectories match the quantum mechanical prediction in position space as well, as can be seen in figure \ref{fig:dslit}(c).
These trajectories intersect, and unlike the de Broglie-Bohm trajectories, they do not necessarily end at the shown screen section.
The initial momenta of the revised trajectories, which are displayed at the bottom of figure \ref{fig:dslit}(d), perfectly match the quantum mechanical probability density \eqref{eq:prob-density-momentum} because they are chosen that way.
At the final time, the histogram is better adapted to the quantum mechanical prediction than the one of the de Broglie-Bohm theory, which is shown in the top of figure \ref{fig:dslit}(b).

However, quantum physics states that the momentum distribution \eqref{eq:prob-density-momentum} is constant in time for the example we are considering.
This is not fulfilled by either the de Broglie-Bohm theory or the revised theory.
The discussion of this discrepancy on the basis of particularly strong deviations, indicated in figures~\ref{fig:dslit}(b,d) by dashed green lines at $t=3.5$\units{ps}, is the subject of the next section.

\section{Conclusion}\label{sec:conclusion}

In section \ref{sec:traj}, we revised the initial momenta of de Broglie-Bohm trajectories to match the quantum mechanical probability density.
However, for the double-slit experiment discussed in section \ref{sec:double-slit}, we could not ensure that the quantum mechanical prediction is fulfilled in momentum space for all times.
This is evident from figure~\ref{fig:dslit}(d), where the momentum distribution should not change with time according to quantum mechanics \eqref{eq:prob-density-momentum}, but the trajectory density does change considerably directly behind the double slit.

\begin{figure}[t!]
	\begin{center}
		\includegraphics[width=15.0cm]{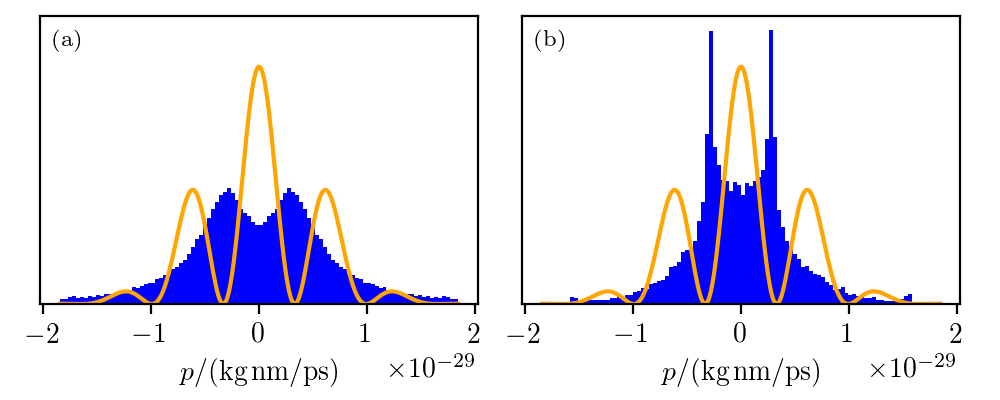}
		\caption{Histogram~(a) shows momentum values which are collected from revised trajectories at the dashed line in figure~\ref{fig:dslit}(d).
        Instead of a maximum in the center of the quantum mechanical probability distribution \eqref{eq:prob-density-momentum} (orange line), the histogram has a minimum.
        Histogram (b), that results from de Broglie-Bohm trajectories at the dashed line in figure~\ref{fig:dslit}(b), displays a similar behavior but with sharper peaks.
        Both histograms contrast strongly with the quantum mechanical prediction.
        This can be explained by a transition from a two-maximum density to a three-maximum density in position space, see figures~\ref{fig:dslit}(a,c).
        Histograms (a,b) could be used to experimentally question trajectory formulations of quantum mechanics.
        }
		\label{fig:comparision}
	\end{center}
\end{figure}

In particular at the dashed line in figure~\ref{fig:dslit}(d), the revised trajectories lead to a momentum distribution that contrasts strongly with the quantum mechanical probability density.
The histogram that results from values at this line is shown in figure~\ref{fig:comparision}(a) and it is compared to the quantum mechanical prediction (orange line).
Instead of a maximum in the center of the orange line, the histogram has a minimum.
Figure~\ref{fig:comparision}(b) shows a similar picture.
The histogram shown there comes from de Broglie-Bohm trajectories at the dashed line in Figure~\ref{fig:dslit}(b).

A possible explanation for these results is provided by examining the corresponding situation around $t=3.5$\units{ps} in position space shown in figures~\ref{fig:dslit}(a,c).
Here, the probability density changes from a two-maximum density to a three-maximum density.
To achieve such a change, the particles need to acquire an increased probability of negative and positive momenta, instead of keeping a high probability for $p$=0.
This explanation accounts for the non-existence of the central peak of the histograms shown in Figure~\ref{fig:comparision}, which is contrary to the quantum mechanical prediction.

It can be assumed that the arguments just given apply not only to the revised theory and to the de Broglie-Bohm theory, but also to other theories where particles are considered to move along well defined trajectories.
This conclusion could be used to question the possibility of trajectory formulations of quantum mechanics by investigating momentum distributions that were measured directly behind a double slit.

\section{Acknowledgment} \label{sec:acknowledgement}

I would like to thank my friends Brian White and Eric Findlay, who have faithfully stood by my side while I worked on the problem discussed here and to whom I owe many valuable suggestions.
I also thank Prof. Jeffrey Zheng for stimulating my interest in the general topic of this manuscript.



\bibliography{dbbrevision}
\bibliographystyle{apsrev4-1}

\end{document}